\documentclass[10pt]{article}

\usepackage{amsmath}
\usepackage{amssymb}

\usepackage{graphicx}

\usepackage{verbatim}

\usepackage{natbib}

\usepackage[labelfont=bf,labelsep=period,justification=raggedright]{caption}

\usepackage[dvipsnames,usenames]{color} 

\newcommand{\bm}[1]{{\bf{#1}}}
\newcommand{\alt}[1]{\lesssim{#1}}
\newcommand{\agt}[1]{\gtrsim{#1}}

\begin{document}

{\title{Epidemics in Networks of Spatially Correlated
Three-dimensional Root Branching Structures }}

\maketitle

\begin{flushleft}

T.~P.~Handford$^{1,\ast}$,
F.~J.~Perez-Reche$^{2}$,
S.~N.~Taraskin$^{3}$,
L.~da~F.~Costa$^{4}$,
M.~Miazaki$^{5}$,
F.~M.~Neri$^{6}$,
C.~A.~Gilligan$^{7}$,
\\
\bf{1}  St. Catharine's College, University of Cambridge,
   Cambridge, UK
\\
\bf{2}  Department of Chemistry, University of Cambridge,
             Cambridge, UK
\\
\bf{3}  St. Catharine's College and Department of Chemistry,
  University of Cambridge,
             Cambridge, UK
\\
\bf{4}  Instituto de Fisica de Sao Carlos,
Universidade de Sao Paulo,
Sao Carlos, SP, Brazil
\\
\bf{5}  Instituto de Fisica de Sao Carlos,
Universidade de Sao Paulo,
Sao Carlos, SP, Brazil
\\
\bf{6}  Department of Plant Sciences, University of Cambridge,
             Cambridge, UK
\\
\bf{7}  Department of Plant Sciences, University of Cambridge,
             Cambridge, UK
\\
$\ast$ E-mail: tph32@cam.ac.uk
\end{flushleft}

\begin{abstract}

Using digitized images of the three-dimensional, branching structures
for root systems of  bean seedlings, together with analytical and
numerical methods that map a common 'SIR' epidemiological model onto
the bond percolation problem, we show how the spatially-correlated
branching structures of plant roots affect transmission efficiencies,
and hence the invasion criterion, for a soil-borne pathogen as it
spreads through ensembles of morphologically complex hosts. We
conclude that the inherent heterogeneities in transmissibilities
arising from correlations in the degrees of overlap between
neighbouring plants, render a population of root systems less
susceptible to epidemic invasion than a corresponding homogeneous
system. Several components of morphological complexity are analysed
that contribute to disorder and heterogeneities in transmissibility of
infection. Anisotropy in root shape is shown to increase resilience to
epidemic invasion, while increasing the degree of branching enhances
the spread of epidemics in the population of roots. Some extension of
the methods for other epidemiological systems are discussed.

\end{abstract}

\section{Introduction}

The identification of efficient strategies is of great practical importance in controlling the spread of epidemics in populations of plants,
animals, social and computer networks
\citep{Nasell_02,Pastor_Satorras_03,Pastor_Satorras_04:book}.
Many previous studies from different disciplines, such as epidemiology,
mathematics, physics, chemistry and social sciences,
have attempted to address this problem using experimental, numerical
and analytical approaches (see e.g.
\citep{Diekmann_00:book,Murray_02:book,Marro_99:book,liggett_85:book,Keeling_07:book,Jackson_08:book}).

Numerous mathematical models exist for the description of
epidemics in populations (see e.g. \citep{Nasell_02} and
references therein).
Typically, a population is represented by a network in which nodes are
associated with hosts and links between nodes describe possible
transmission paths for diseases.
The hosts can be in several
states such as susceptible (S),
infected (I) and recovered (or equivalently removed) (R).
In two prototype epidemiological models, SIS and SIR, the hosts change
their state according to certain rules.
For example, in the SIS model, a susceptible host (S) can be infected by
its neighbour (I) and then again become susceptible, thus following
the sequence, S$\to $I$\to$S.
The SIR model differs from the SIS only in the last step of the
sequence, S$\to $I$\to$R, at which an infected host
is effectively removed from the epidemic either by death or by becoming immune to further infection.
Extensions and modifications of these two models have been proposed to
describe  various features of realistic epidemics (see
e.g. \citep{Dybiec_04}).

Both models follow stochastic dynamical rules meaning that either
one or both of the dynamical processes, infection and/or recovery, occur
randomly thus reflecting
the inherent complex nature of the hosts
and communications between them.
Depending on the relative value of the infection and recovery rates,
the epidemic can be in different regimes: non-active (epidemic does not
cover the major part of the population) and active (outbreak of
epidemic).
The transition between these two regimes occurs
at a critical point characterized by the critical value of a control
parameter
(e.g. the ratio of
recovery and infection rates) and can be described as a continuous
phase transition belonging to the universality class of
directed or isotropic percolation for SIS and SIR models,
respectively
\citep{Marro_99:book,Hinrichsen_00:review,Odor_04:review}.
Universality at the threshold refers to universal critical exponents
describing  scaling behaviour of e.g. correlation length in spatial
fluctuations of hosts in a certain state.
In contrast, the threshold value of the control parameter is
system-dependent and can vary significantly between models
belonging to the same universality class.

The location of the threshold in the parameter space and
  the evaluation of the probability of invasion above the threshold
  provide information about the vulnerability
  of the system to epidemic outbreaks. The value for the
  threshold is known
  analytically only for some
  simple models: for example the threshold is known for both the SIR and SIS
models in the case of networks with homogeneous transmission rates and given degree
distributions \citep{Pastor_Satorras_03}). More usually, however,
numerical methods are required to identify both the probability and
threshold for invasion. The derivation of analytical insight is
further challenged for many realistic systems that are characterized
by inherently heterogeneous parameters which may also be correlated in
space.

Heterogeneity can occur amongst the nodes and connections in epidemiological networks. Thus individual host characteristics, such as host type and age, recovery time,
  infectivity and immunity may differ; there may also be variation in the transmissibility of infection between pairs of infected and susceptible hosts.
\cite{PerezReche2010}, recently considered heterogeneity amongst nodes by analysing the influence of host morphology on the properties of epidemics in a set of two-dimensional (2d) retinal ganglion cells placed on the nodes of a regular 2d lattice.
They  showed that heterogeneity associated with host morphology significantly
influenced the invasion threshold, making a system of morphologically
complex hosts less vulnerable 
 to epidemic invasion \citep{PerezReche2010}.
\cite{PerezReche2010} also showed that, although the set of neural cells exhibited heterogeneity, there was little evidence for correlation in the transmission of infection between pairs of contiguous hosts. Transmission could therefore be successfully
described \citep{PerezReche2010} by an equivalent effective mean-field
homogeneous system \citep{Sander_02}. Many systems, however, exhibit
greater heterogeneity than the 2d ganglion system especially in the
degree of anisotropy and in transmissibility between hosts. In this
paper, we advance the analyses to 3d morphologically complex and
anisotropic  hosts (plant root systems) and demonstrate that high
anisotropy in 3d root systems can bring non-negligible correlations in
pair-wise transmission of pathogens from infected to susceptible roots
systems. Such correlations lead to a break-down in the mean-field
description and make these systems 
safer to epidemic
outbreaks.

Below we focus on the spread of soil-borne disease through
populations of plants in which the transmission of infection occurs between
contiguous plants and is mediated by the complex morphology of
branching root structures \citep{Gilligan_1994}.
Specifically the aims of the paper are (i) to study analytically and
numerically the spread of epidemics in a population of plants with
realistically complex 3d morphology,(ii) to locate the associated
threshold for invasion and subsequent epidemic spread,
(iii) to predict how the invasion threshold depends
upon the shape of the 3d hosts and
(iv) thus to suggest how to
control the vulnerability 
 of the system to pathogen invasion by
changing the host morphology.
In this paper, we consider a population of plants, each comprising
an ensemble of roots, with the plants arranged on a regular
two-dimensional lattice to mimic the spatial arrangement of
plants in a field. For concreteness, we deal with an SIR model in which
  infected plants  eventually die (i.e. permanently go to the R-class). The SIR model can be applied to a wide range of soil-borne
  pathogens \citep{Hornby_90:book}  spreading through populations of
  roots \citep{Gilligan_1990, Gilligan_2002}.
In what follows, we
 analyse how the morphology of the root systems of bean
  (\emph{Phaseolus vulgaris}) seedlings, affect the probability of
  epidemic invasion of a common and ubiquitous soil-borne, fungal
  pathogen, \emph{Rhizoctonia solani}, which spreads by mycelial
  extension from infected to susceptible hosts
\citep{Bailey_2000}.

Our main findings are the following.
(i) The morphology of the hosts represented by 3d roots is an important
factor for the SIR epidemic.
Two morphological features, namely anisotropy and disorder, are of particular
importance for the spread of epidemics.
Anisotropy refers mainly to the deviation of the major root stems from the
vertical direction.
Disorder is a characteristic related to the degree of root branching
and in particular to the number of secondary branches in the root.
(ii) Anisotropy brings short-range  correlations to the disease
transmission which makes the system less vulnerable to epidemic
  invasion.   
(iii) Disorder diminishes the effect of anisotropy and thus 
increases the vulnerability of the system to an epidemic outbreak.

The structure of the paper is as follows.
In Sec.~\ref{sec:model}, the SIR model is introduced for the root
system placed on a square lattice.
The main results for invasion probability and analysis of
correlations in transmissibilities  are presented in
Sec.~\ref{sec:results}.
Discussion and conclusions are given in Sec.~\ref{sec:discussion}.


\section{Model
\label{sec:model}}

\subsection{SIR process
\label{subsec:SIR}}

In the SIR model used below, the hosts
can be in one of the three states, susceptible~(S-class), infected~(I-class) or
recovered and \emph{permanently} immune to infection (R-class).
Here a host refers to a plant root system. Each plant is
located on the node of a square (for concreteness) lattice
with links between nearest-neighbours only. 
An infected host remains in such a state
for a fixed time $\tau$ and then ceases to be
infectious (i.e. permanently goes to the R-class).
The recovery time is chosen to be the time scale for the problem and set
to $\tau=1$.
During the infectious period, while times are in the range $t
  \alt \tau$, an infected node can
stochastically, by a Poisson process, transmit
the pathogen and hence disease to its nearest neighbour with
a rate $\beta$.
Under these rules, the probability, $T$, that disease is passed from
an infected host to its susceptible neighbour before that
  host moves to the R~state is given by the following
expression~\citep{Grassberger_83},
\begin{equation}
T=1-e^{-\tau\beta}\equiv 1-e^{-\beta}~.
\label{eq:transmissibility}
\end{equation}
This value, called the transmissibility, quantifies the
  transmission process between neighbouring hosts. 
At a large scale, $T$ determines the probability $P_{\text{inv}}$
  that a large outbreak occurs, with the pathogen eventually invading
  the system (invasion probability).  
The probability $P_{\text{inv}}$ is difficult to calculate
  analytically, 
and typically it is computed numerically by generating
many stochastic realizations of epidemics and counting the fraction of
those that span the system across all borders.

In the language of phase transitions, $T$ is the control 
parameter for the homogeneous SIR model (it is varied to study
the
transition) and $P_{\text{inv}}$ is the order parameter. 
If transmissibility is less than some critical value, $T<T_c$,
then epidemic invasion cannot 
occur in
an
infinite system and
the invasion probability for the epidemic
is zero,  $P_{\text{inv}}=0$.
If  $T>T_c$, the invasion probability is finite,  $P_{\text{inv}}>0$,
approaching unity when $T\to 1$.
The critical value of transmissibility coincides with
the critical bond probability, $p_c$, for the bond percolation problem
 because
the final cluster of R sites in the SIR
problem coincides with one of the clusters of sites connected by
occupied bonds in the bond percolation
problem, if the bond probability is equal to
the transmissibility~\citep{Grassberger_83}.
In particular, $T_c=p_c =1/2$  for a square lattice with homogeneous
transmissibilities  \citep{Isichenko_92:review,Stauffer_92:book}.

An epidemic with heterogeneous uncorrelated transmission rates (for fixed value of $\tau$) can also be
mapped onto the bond-percolation problem, so that at criticality the
following condition should hold
\citep{Sander_02,Newman_02:epidemic,Kenah_07,Miller2007},
\begin{equation}
\langle T \rangle =\int\limits_0^\infty
(1-e^{-\tau\beta})\rho(\beta)\text{d}\beta = p_c~,
\label{eq:criticality_mf}
\end{equation}
where $\rho(\beta)$ is the probability density function for
  the transmission rate $\beta$ and $p_c$ is the bond percolation threshold for
the lattice on which
the SIR model is defined.
Moreover, this mean-field relation is valid for the whole range of
transmissibilities and the invasion probability in
a heterogeneous system can be shown to be (\cite{Sander_02}) 
\begin{equation}
P_{\text{inv}}(\{T_{ij}\}) = P_{\text{inv}}(\langle T \rangle)~,
\label{eq:P_inv_mf}
\end{equation}
 i.e. an epidemic in an uncorrelated
heterogeneous system with transmissibilities
 $\{T_{ij}\}$ is equivalent to
an epidemic in a
 homogeneous system in which all the transmissibilities are replaced
by the mean value of transmissibility $\langle T \rangle$.

The above statement is correct only in the case of independent
transmissibilities between different pairs of hosts.
Below, we use Monte-Carlo simulations of multiple possible
  invasion events to demonstrate that for realistic hosts with
complex morphology
this assumption does not necessarily hold and a simple mean-field
description of the SIR process can fail.
Namely, the inherent short-range correlations in transmissibilities between
different pairs of hosts are shown to change the threshold value for
transmissibility.

\subsection{Root systems as particular hosts
\label{subsec:roots}}

\begin{figure}[htp]
\includegraphics[scale=1]{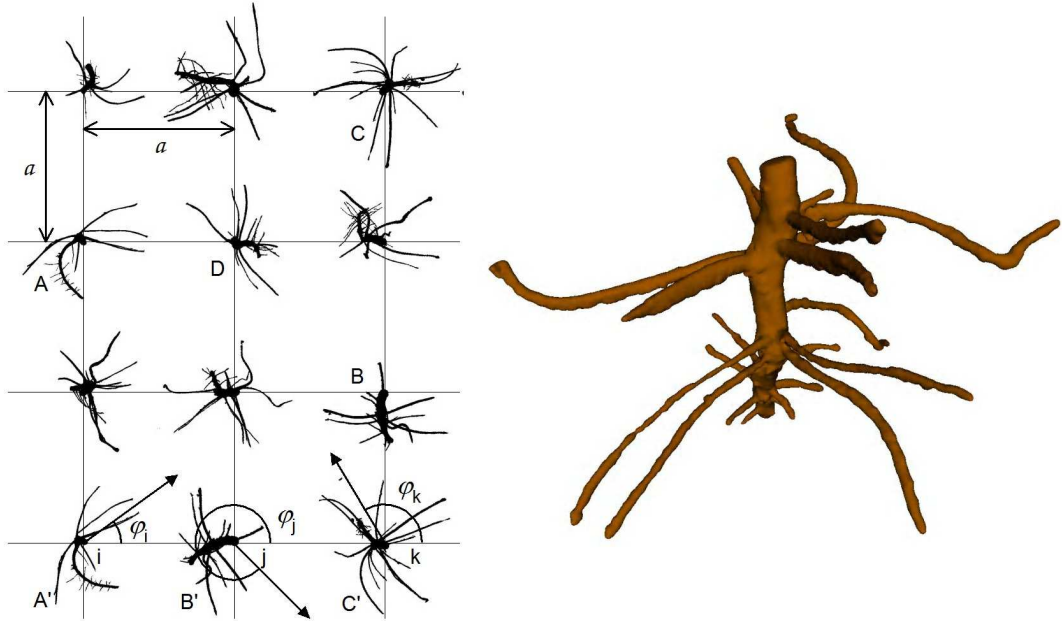}
\caption{\label{fig:rootProj}
Projections onto a horizontal
plane of the root systems of nine bean plants placed on a square
lattice with spacing $a$  are shown in the first three rows of the
left panel. The last row shows three root systems A', B' and C' which
are obtained by rotation about  a
vertical axis
of randomly chosen plants A, B and C
by random angles $\varphi_i$, $\varphi_j$ and $\varphi_k$, respectively.
This row represents a typical pattern of the lattice model studied here.
The right panel shows a 3 dimensional image of
the root system D from the left panel.
}
\end{figure}

Realistic
bean root systems were used as hosts in our model.  Nine bean
seeds, $N_r=9$, were allowed to germinate for 2 days on cotton, and
then transferred to an aquarium to grow hydroponically
over 4 days, before being set into paraffin blocks which
were then sliced into 0.1 mm thick layers.
The layers were digitally scanned with pixels of a linear size
equal to $0.1$~mm, 
  to produce three-dimensional images of the
primary (tap) root and first-order lateral roots, hereafter
referred to as \emph{root systems} (see a typical image in the right
panel of Fig.~\ref{fig:rootProj}).

\vskip20pt

The root systems thus obtained (see Fig.~\ref{fig:rootProj})
 were used as hosts in a lattice model to study the invasion of a
 soil-borne SIR epidemic spreading through a population of plants by
 mycelial growth from root system to root system.
The lattice model was created in the following manner.
An arbitrary root system (e.g. A in Fig.~\ref{fig:rootProj})
was selected from the nine experimentally available
systems (see first three rows in the left panel of
Fig.~\ref{fig:rootProj}).
This root system was
rotated by a random angle $0 <\varphi< 2\pi$ about a vertical axis
passing through the original bean and
then placed on a node (with the bean-seed position coinciding with the node) $i$
($i=1,2,\ldots,L^2$) of
a square lattice of size $L \times L$ with lattice spacing $a$
(e.g. root A' in Fig.~\ref{fig:rootProj}).
 This process was repeated for all the nodes of the lattice.
 Several realizations ($\sim 10^3$) of so created lattice systems form a
 statistical ensemble which has been analysed below.

A soil-borne epidemic can be modelled by introducing a pathogen, in
the form of a small fungal colony typical of \emph{R. solani} onto a
root system near the centre of the lattice. The pathogen is assumed to
infect the initial plant and to grow along and around the root system
by mycelial extension exploring the soil nearby.
If another root system is situated nearby,
the fungal colony can
reach it and eventually infect it.
In this way, the fungal colony propagates through a
population of roots.

The transmission rate between plants in the SIR process, which is
effected by mycelial spread between root systems, is
proportional to the effective
overlap between different 3d root systems.
In order to characterize this overlap quantitatively, we first define
the root number density (number of voxels per volume) in the
following way.
The roots (comprising the primary root and first-order lateral
  roots) in our model are represented by 3d digital images.
Each image, $i$, consists of a set of $N_i$ voxels
(three-dimensional  pixels) with volume
$V_{\text{voxel}}=10^{-3}\,\text{mm}^3$, the locations of
which are given by position vectors
$\bm{r}_{ni}=(x_{ni},y_{ni},z_{ni})$ with $n=1,\ldots ,N_i$.
The root number density that the image
represents, $\rho_{\text{scan},i}(\bm{r})$,  is equal to
$1/V_{\text{voxel}}$ if $\bm{r}$ coincides with the position of one of
the voxels and zero otherwise.
On the scale of the system, the number density of voxels can be
  written as a sum of Dirac  delta-functions,
\begin{equation}
\rho_{\text{scan},i}(\bm{r})=\sum_{n=1}^{N_i}\delta(\bm{r}-\bm{r}_{ni})~.
\label{eq:inpDens}
\end{equation}
Bearing in mind that the density should describe some additional
  volume around each root across which the fungus can spread though
  soil, known as the pathozone \citep{Gilligan_95}, and which therefore
  mediates the transmission of infection from a donor (i.e. infected)
  to a recipient (susceptible) root system, we apply broadening to
the images replacing the Dirac delta-functions
in Eq.~(\ref{eq:inpDens}) (or voxels)
by their Gaussian representation, i.e.
\begin{equation}
\rho_i(\bm{r})\simeq \int
\text{d}\bm{r'}\rho_{\text{scan},i}(\bm{r'})\left({1\over{2\pi\sigma^2}}
\right)^{3/2}\exp{\left[-{(\bm{r}-\bm{r'})^2\over{2\sigma^2}}\right]}~,
\label{eq:broaden}
\end{equation}
with the broadening parameter $\sigma = 1.5$~ (the value of $\sigma$
is much smaller than the lattice spacing, i.e. $\sigma \ll a$, so that
the overlap with next-nearest neighbours and thus 
transmission of the pathogen 
to them can be ignored in the model) of the
 linear size of the  voxel 
(pixel) (the abbreviation {\it px} is used below for pixel).
This mimics the diffusive spread of a pathogen into the medium
surrounding the root \citep{Gilligan_95}.

The fungal colony can efficiently pass between different
roots through the volume of soil if branches of both roots are
present in that volume, i.e. if the roots overlap.
Therefore, we assume that transmission between nearest-neighbour
 hosts $i$ and $j$ separated by unit cell vector
takes place at a rate
$\beta_{ij}$, proportional to the 3d overlap $J_{ij}$ between the two
hosts, i.e

\begin{eqnarray}\label{eq:beta}
\beta_{ij} = kJ_{ij}~,
\end{eqnarray}
where,
\begin{eqnarray}
J_{ij}&=&V_{\text{voxel}}
\int\text{d}\bm{r}\rho_i(\bm{r})\rho_j(\bm{r})
\nonumber
\\
&=& V_{\text{voxel}}
\sum_{m=1}^{N_i}\sum_{n=1}^{N_j}\left({1\over{4\pi\sigma^2}}\right)^{3/2}
\exp{\left[-{(\bm{r}_{mi}-\bm{r}_{nj})^2\over{4\sigma^2}}\right]}~
~.
\label{eq:Noverlap}
\end{eqnarray}
The value of $J_{ij}$ (dimensionless) is the number of voxels of root
$i$ in the region
of the overlap with root $j$. From the definition, $J_{ij}=J_{ji}$.
The coefficient $k$, the infection efficiency, has the meaning of
transmission rate per overlapping voxel and is set to be the same
for all overlapping roots.
 This efficiency represents the ability
  of the pathogen to utilise a point of contact between two hosts to
  spread between them.
 It therefore takes into account all 
  details
 of the transmission process not associated with the host morphology
 and specific to the particular host-pathogen system, such as 
 susceptibility and
 infectivity, together with other factors such as environmental
 conditions 
that, in turn, influence transmission of infection.
 We therefore investigate the role of morphology in deciding the
 overall rate between two root systems, with all other properties
 decribed by variable values of $k$.
Therefore, both the transmission rate, $\beta_{ij}$,  between an
arbitrary pair of
the nearest-neighbour hosts and thus transmissibility,
\begin{eqnarray}
T_{ij}=1-\exp\{-kJ_{ij}\}
~,
\label{eq:Tij}
\end{eqnarray}
 (see Eq.~(\ref{eq:transmissibility})),  depend
on the transmission efficiency
$k$, being a control parameter, and  the
random overlap  $J_{ij}$
which depends on distance between hosts $a$, being another control
parameter.
The randomness in the
overlaps is caused by the complex morphology of hosts.
With these assumptions, all the information about a given spatial configuration of the system (a particular realization of roots and angles at each node) is contained in the set of overlaps $\{J_{ij}\}$, while all the information about the epidemiological properties of the system is contained in the set of transmissibilities $\{T_{ij}\}$.

\begin{figure}[tp]
\includegraphics[scale=0.6]{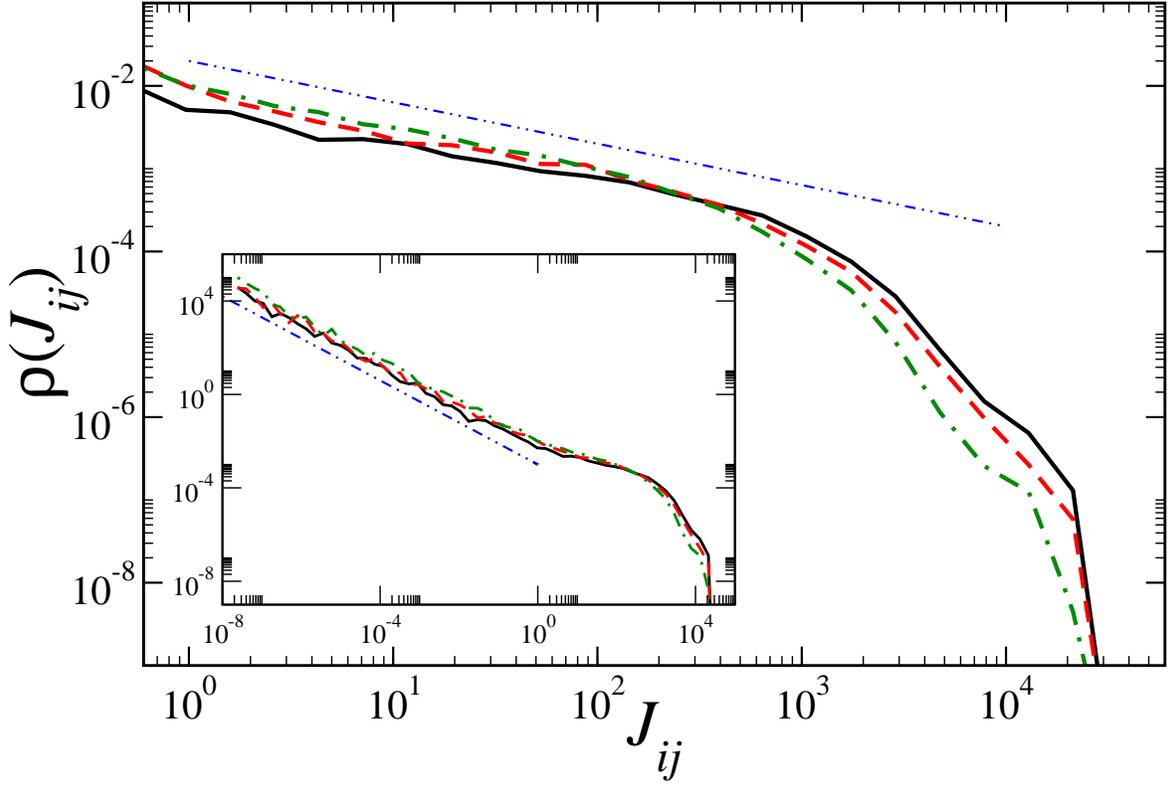}
\caption{\label{fig:rho_J}
Probability density function $\rho(J_{ij})$ for
overlaps, $J_{ij}$, between root systems
in double-log scale
for different lattice spacings: $a=100$ px$^{1/2}$ (solid curve),
$a=120$  px$^{1/2}$ (dashed) and $a=150$  px$^{1/2}$ (dot-dashed).
The double dot-dashed line corresponds to $\rho(J_{ij})\propto
J_{ij}^{-\alpha}$ with $\alpha=1/2$ (guide for eye only).
The inset shows the same distributions on a larger scale with the
 double dot-dashed line corresponding to $\rho(J_{ij})\propto
J_{ij}^{-\alpha}$ with $\alpha=0.9$ (guide for eye only).
}
\end{figure}

\begin{figure}[hp]
\includegraphics[scale=0.6]{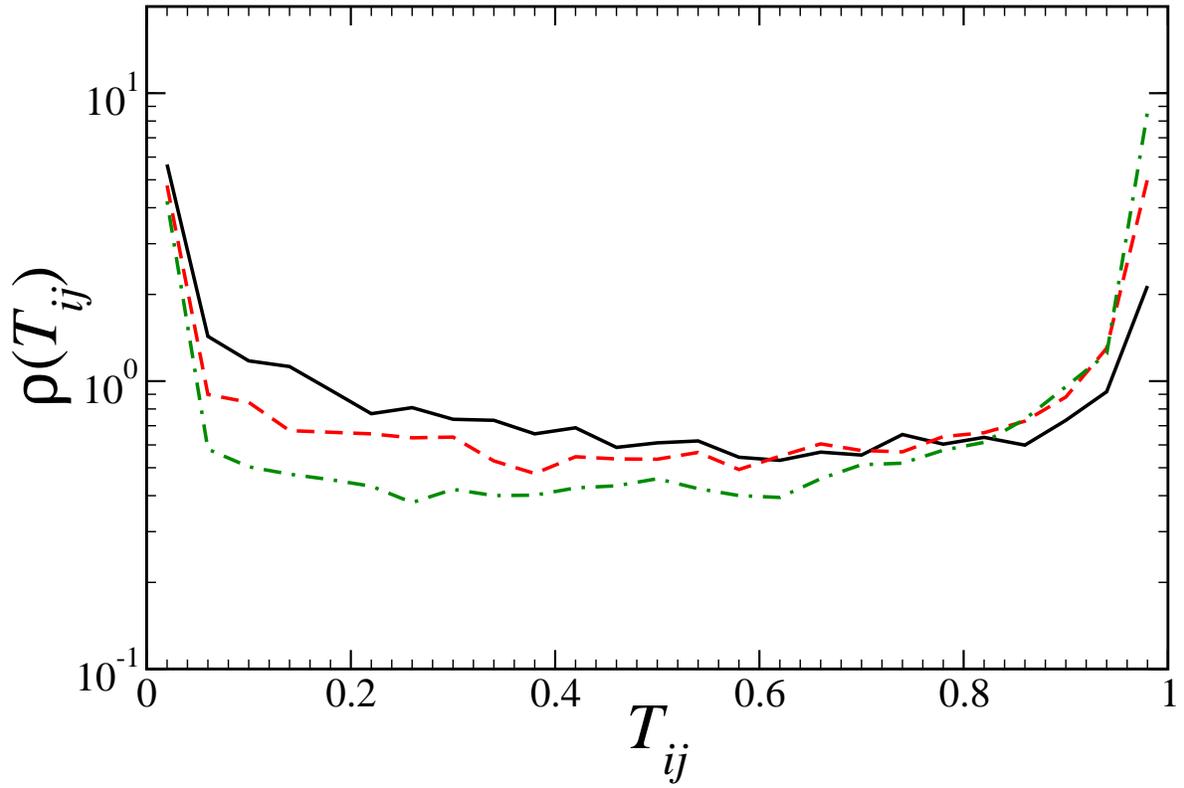}
\caption{\label{fig:rho_T}
Probability density function $\rho(T_{ij})$ for transmissibilities, $T_{ij}$,
in semi-log scale for three different values of
the transmission efficiency, $k$: $k=k_c/2$ (solid curve), $k=k_c$ (dashed
curve) and $k=2k_c$ with $k_c \simeq 0.00344$ for $a=120$ px$^{1/2}$.
}
\end{figure}

Typical probability densities $\rho(J_{ij})$ and $\rho(T_{ij})$ for
the ensemble of roots  are
shown in Fig.~\ref{fig:rho_J} and  Fig.~\ref{fig:rho_T}.
It follows from Fig.~\ref{fig:rho_J} that $\rho(J_{ij})$ is a
monotonically decaying function.
The value of $J$ roughly corresponds to
the number of pixels in the
overlap region.
In the region of very small overlaps $J_{ij}\alt 1$
(i.e. the overlap is either less than or of order of unity)
caused by
Gaussian broadening effects for pixels,
\begin{eqnarray}
\rho(J_{ij})\propto J_{ij}^{-\alpha}
 ~,
\label{eq:rho_J}
\end{eqnarray}
with $\alpha \simeq 0.9$ (see the inset in Fig.~\ref{fig:rho_J}).
For the most interesting range of $1\alt J_{ij}\alt 10^3$, the probability
density decays according to Eq.~(\ref{eq:rho_J}) with $\alpha \simeq
0.5$ (see Fig.~\ref{fig:rho_J}).
For larger values of $J_{ij}$ the function $\rho(J_{ij})$ exhibits even
  faster decay.

The shape of the probability density for transmissibilities follows
from  the mapping given by Eq.(\ref{eq:Tij}).
The semi-infinite range of $J_{ij}$ is mapped onto  a finite interval
for transmissibilities, $0\le T_{ij}\le 1$, in such a way that all the large
values of $J_{ij}$ are accumulated around $T_{ij}\simeq 1$ and all
small values of  $J_{ij}$ are concentrated around  $T_{ij}\simeq 0$.
As a consequence the monotonic distribution $\rho(J_{ij})$ gives rise to a
distribution with two maxima (see
Fig.~\ref{fig:rho_T})).
The relative weights of the peaks depend upon
the value of  the transmission
efficiency, $k$.
For small values of $k \alt k_c$, the peak around zero is dominant
(see the solid line in Fig.~(\ref{fig:rho_T}))
ensuring the non-active regime for epidemics, i.e. there is no invasion.
In contrast,
for large values of $k\agt k_c$, the peak around unity
becomes dominant for the
distribution of transmissibilities and the SIR
process is active
(see the dot-dashed line in Fig.~(\ref{fig:rho_T})).
At criticality, $k = k_c$,  the distribution of transmissibilities is
approximately symmetric around $T=1/2$
(see the dashed line in Fig.~(\ref{fig:rho_T})).

\vskip20pt

\vskip20pt

\section{\label{sec:results}Results}

In this section, we present our main results on the behaviour of the SIR
epidemic for the ensemble of
 realistic 3d root systems of bean plants placed on a regular
lattice.
We start with an evaluation of the invasion probability (order parameter)
as a function of two control parameters, transmission efficiency and
lattice spacing, paying special attention to
 the threshold values of the control parameters separating
 invasive and non-invasive regimes of epidemics. 
After that we compare the behaviour of the epidemic in our model with
a mean-field system and find a significant role of correlations in
transmissibilities of pairs of bonds separated by a relatively
short distance.
Finally, several ``toy'' models are investigated in order to
clarify the effect of correlations.

\subsection{\label{subsec:phasediagram} Phase-diagram}

One of the most important characteristics for epidemic spread is the
probability of invasion.
In the language of continuous phase transitions, the invasion
probability, $P_{\text{inv}}$, is the order parameter defining the
status of the epidemic to be either in  the non-invasive
  ($P_{\text{inv}}=0$) or  invasive ($P_{\text{inv}}>0$) regimes.
In our model, the invasion probability depends on two control
parameters, the transmission efficiency (specific to the host-pathogen system and independent of the spatial configuration) and the lattice spacing (affecting the set of overlaps $\{J_{ij}\}$),
$P_{\text{inv}}=P_{\text{inv}}(k,a)$.

In order to calculate $P_{inv}(k,a)$ for a particular set of transmissibilities
$\{T_{ij}\}$, we use the mapping to bond percolation
and create many stochastic realizations in which each bond $i-j$
is occupied with probability $T_{ij}$.
The probability of invasion is defined in an infinite system as
  the probability that any given site belongs to an infinite, or spanning,
  cluster of sites.
 A single such spanning cluster will always exist
  above the invasion threshold in an infinite system, and will never
  exist below it.
 Therefore the invasion probability is equal to the
  fraction of sites out of the whole system that  belong to this
  cluster, when it exists, and zero otherwise.
 Since it is clear that there will not be an infinite cluster in a
 finite system (which is all we can model), we define a spanning cluster
 here as one which spans the lattice touching all four boundaries.
Then for each stochastic realization of the bonds, we determine if
the spanning cluster exists, and what fraction of the
system it makes up, thus obtaining an estimate of $P_{\text{inv}}$ from
that realization.
The average of these estimates 
over all stochastic realizations gives the probability of invasion for the
particular values of $\{T_{ij}\}$.

A different realization of the set $\{T_{ij}\}$ would produce a slightly
different value of the invasion probability.
Therefore we have averaged the invasion probability over different
realizations of $\{T_{ij}\}$, to obtain
the \emph{configurational average} of the
invasion probability $\langle P_{inv}(k,a) \rangle$.
This value is representative because the invasion probability is a
self-averaging quantity \citep{Binder_86} as we have checked
numerically.

\vskip20pt
\begin{figure}[p]
\includegraphics[angle=270,scale=0.6]{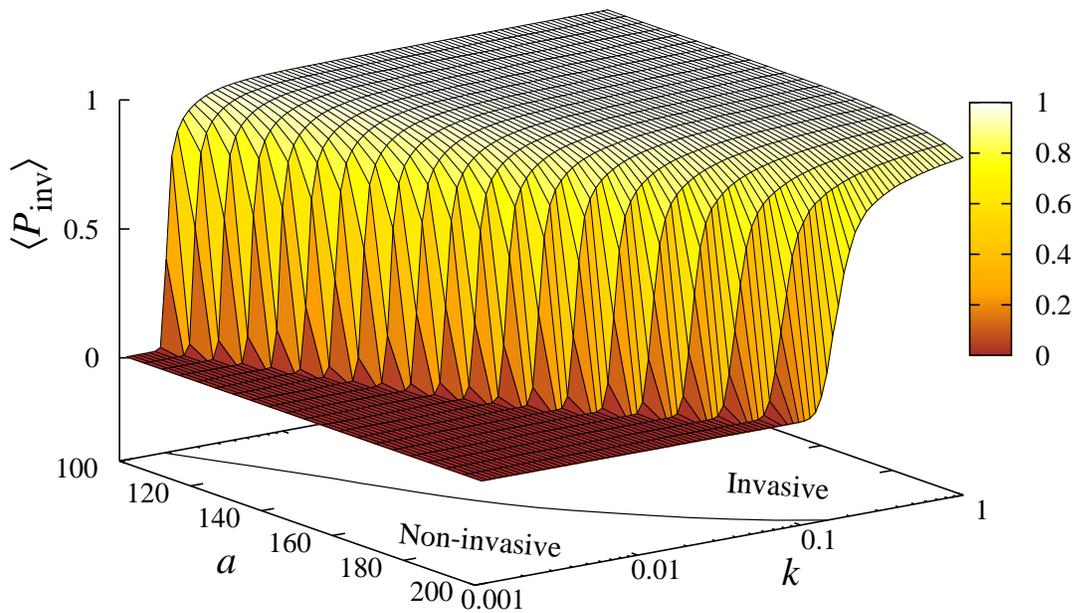}
\caption{\label{fig:3d_phase_diagram}
The dependence of the configurationally averaged invasion probability,
$\langle P_{inv}(k,a) \rangle$, on  the transmission efficiency, $k$
(log-scale),  and lattice spacing, $a$.
The averaging has been done over $1000$ configurations of root systems on
200$\times$200 lattice with open boundary conditions.
The line in the $a-\log(k)$ plane represents the phase boundary for
the invasive and non-invasive regimes of an SIR epidemic.
The critical values of $k$ for given $a$ have been obtained by
finite-size scaling collapse with relative error $<2\%$.
}
\end{figure}

The results of numerical simulations for the average invasion
  probability, $\langle P_{inv}(k,a;L)
\rangle$, for a finite size, $L\times L$, lattice
are presented in Fig.~\ref{fig:3d_phase_diagram}, with $L=200$.
As expected, for large values of lattice spacing and small transmission
efficiency the system is unsuited to pathogen invasion.
In contrast, small lattice spacings and high values of $k$ enhance the
probability of invasion and hence of an epidemic.

The invasion probability surface $\langle P_{inv}(k,a;L) \rangle$
shown in Fig.~\ref{fig:3d_phase_diagram} depends on the lattice size $L$.
When the system size becomes very
large, $L\to \infty$,
the values of the probability of invasion tend to a certain limiting
surface.
The line of the critical points $a_c = a_c(k_c)$ (the solid line
in the horizontal plane in Fig.~\ref{fig:3d_phase_diagram}) separating
non-invasive
($\lim_{L\to\infty}\langle P_{inv}(k,a|L) \rangle =0$) and invasive
($\lim_{L\to\infty}\langle P_{inv}(k,a|L) \rangle >0$)
regimes is of particular importance because it allows the safe
(non-invasive) parameter region to be identified.

The data points on the phase boundary  $a_c = a_c(k_c)$  were obtained by
finite-size scaling collapse in the following way.
The dependence of the order parameter on the system size around the critical point (e.g. $k_c$) for continuous phase transition is well established
\citep{Isichenko_92:review},
\begin{equation}
\langle P_{\text{inv}}(k,a;L) \rangle =
L^{-\frac{\beta}{\nu}}
\tilde{P}_{\text{inv}}((k-k_c)L^{\frac{1}{\nu}})~,
\label{eq:scaling}
\end{equation}
where $\tilde{P}_{\text{inv}}$ is the scaling function and $\beta$ and
$\nu$ are the universal scaling exponents.
The scaling function can be found by scaling collapse of $\langle
P_{\text{inv}}(k,a;L) \rangle$ for several system sizes using $k_c$
and scaling exponents as fitting parameters
\citep{Heermann_80}. Technically this is achieved by plotting
  $\langle P_{\text{inv}}(k,a;L) \rangle L^{\frac{\beta}{\nu}}$ versus
  $(k-k_c)L^{\frac{1}{\nu}}$ for several values of $L$, and finding
  the values of $\nu$, $\beta$ and $k_c$ that lead to a collapse of all
  the curves for different $L$ to a single one, which corresponds to
  $\tilde{P}_{\text{inv}}$.
Such a finite-size scaling analysis was applied to evaluate the
critical values of $k_c$ for given lattice spacings and the results
are shown by the line in the horizontal plane in
Fig.~\ref{fig:3d_phase_diagram}.
The scaling collapse also gives the values of the scaling exponents,
$\beta = 0.13 \pm 0.008$ and $\nu = 1.32 \pm 0.05$,
which  coincide with those for the isotropic
percolation universality class \citep{Isichenko_92:review}.
We have also performed the scaling collapse for the susceptibility
\citep{Heermann_80} and found
the expected values  for the corresponding scaling exponents
and the invasion threshold.
Therefore, we have confirmed that the continuous phase transition for
invasion probability in the system of roots (with short-range
correlations discussed below), as expected, is similar to that in
isotropic percolation, i.e. it belongs to the isotropic
percolation universality class.

\subsection{\label{subsec:Correlations}Correlations}

\begin{figure}[htp]
\includegraphics[angle=0,scale=0.6]{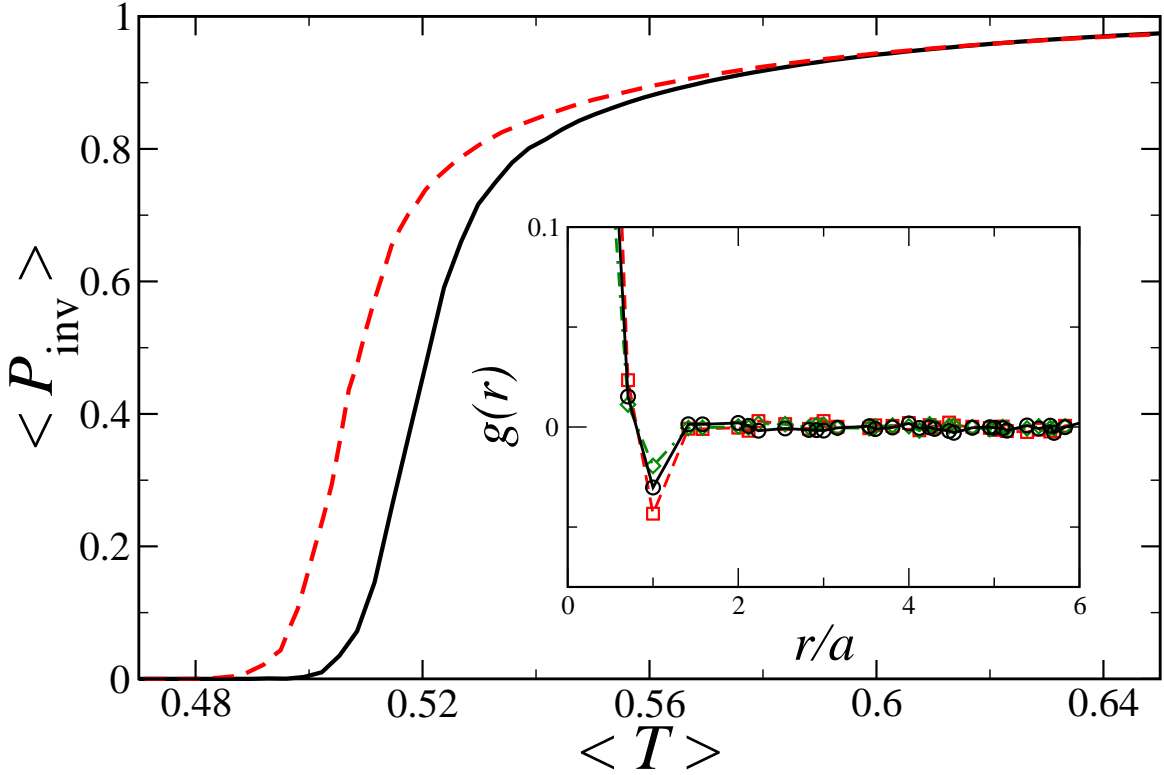}
\caption{\label{fig:PvsT}
Invasion probability $P_{\text{inv}}$ {\it versus} mean transmissibility
$\langle T\rangle$ for the realistically complex roots systems in which
transmissibilities between two hosts are calculated according to
Eq.~(\ref{eq:Tij}) (the solid curve) and the mean-field system in which
all the transmissibilities coincide with the mean transmissibility
$\langle T \rangle$ (the dashed curve).
The lattice spacing is  $a=120$~px$^{1/2}$ and values of all other parameters
are the same as those used in Fig.~\ref{fig:3d_phase_diagram}.
The inset shows the spatial correlation function, $g(r)$, for
transmissibilities evaluated at critical transmission efficiency,
  $k_c(a=120) \simeq 0.00344$ (circles connected by solid line),
$k=k_c/2$ (squares connected by the dashed line) and $k=2k_c$
(diamonds connected by the dot-dashed line).
}
\end{figure}

The invasion probability presented in Fig.~\ref{fig:3d_phase_diagram}
for a heterogeneous lattice with transmissibilities $\{T_{ij}\}$
has been calculated numerically by
 estimating the mean fraction of sites belonging to the spanning
cluster in the bond-percolation problem.
Alternatively, we could use a mean-field assumption
  (Eq.~(\ref{eq:P_inv_mf})), which suggests that the heterogeneous
system is equivalent to a homogeneous one with average
transmissibility $\langle T \rangle$, if there are no
correlations between transmissibilities. This would prove that the spread of the soil-borne pathogen in the root system is not be affected by the heterogeneity of the hosts.
To test the validity of the mean-field approach for our model, in
Fig.~\ref{fig:PvsT}, we show the dependence of the invasion
probability on mean transmissibility both for heterogeneous
(solid curve) and homogeneous mean-field (dashed curve)
systems.
Evidently,  the critical value of the transmissibility in the real-root
system is greater than that in the mean-field system,
and thus the real-root system with inherent heterogeneity in
transmissibilities is less vulnerable 
 to epidemic invasion as compared
with the homogeneous system with the same MEAN transmissibility.
This implies the presence of correlations in transmissibilities
  between hosts in the system \citep{Sander_02,PerezReche2010}. Note
  that these correlations are of a
  different nature to spatial-temporal correlations in the density of
  infected/removed hosts observed during the propagation of disease.
The short-range correlations in transmissibilities make an epidemic
less likely in the heterogeneous system under
consideration (solid curve in Fig.~\ref{fig:PvsT}) than in a
homogeneous mean-field system (dashed curve)
for values of  $P_{\text{inv}} \alt 0.95$.

It should be noted that for relatively large values of the mean
transmissibility (e.g. $\langle T \rangle \agt 0.62$ in
Fig.~\ref{fig:PvsT}),
the invasion probability for
heterogeneous system is higher than for homogeneous one. This is a
consequence of negative correlations in transmission
(cf. Fig.~\ref{fig:PvsT}).  This behaviour is in contrast with
that for systems with non-negative correlations in transmission such
as, e.g. a system with heterogeneous recovery times. Indeed, in this
case it can be rigorously shown that the probability of invasion in
a homogeneous system cannot be smaller than that for heterogeneous
systems for any value of $\langle T \rangle$
\citep{Kuulasmaa_82,Cox_88,Kenah_07,Miller2007}. However, this theorem does not apply to
systems with negative correlations, such as the root system studied
here.
%
%
%
%

The difference in the invasion curves for heterogeneous and
homogeneous systems (cf. solid and dashed curves in
Fig.~\ref{fig:PvsT})  is a clear indication of the correlations in
transmissibilities for heterogeneous systems.
Indeed, the spatial correlation function for transmissibilities,
\begin{equation}
g(r)=\frac{\langle T_{ij}T_{km}\rangle-\langle T_{ij}\rangle\langle
  T_{km}\rangle}{\langle T_{ij}\rangle\langle T_{km}\rangle} =
\frac{\langle T_{ij}T_{km}\rangle}{\langle T\rangle^2}
-1 ~,
\label{eq:c_of_r}
\end{equation}
plotted in the inset to Fig.~\ref{fig:PvsT}, shows the presence of
negative short-range (between nearest bonds) correlations (see the dip
at $r/a=1$ in the inset of Fig.~\ref{fig:PvsT}).
In Eq.~(\ref{eq:c_of_r}), the distance $r$ is taken between midpoints
of different bonds $i-j$ and $k-m$ and the averaging is performed over all
bond pairs separated by the same distance.

Therefore, the SIR process in the system of real roots can be mapped
onto the bond-percolation problem with short-range correlations between bond
probabilities. In the next subsection, we reveal the origin of such correlations.

\subsection{Disc-Based Models
\label{subsec:Models}}

In order to understand the origin of the short-range correlations in
transmissibilities between roots, we  study
 three simplified ``toy'' models of differing complexity.
In all these models, each
root system $i$ is replaced  by a two-dimensional horizontal ``disc'',
i.e. by the
number density $\rho_i({\bm r})$ with the centre of mass at ${\bm
  r}_i$.
The centres of the discs are randomly displaced from the lattice nodes
characterized by position vectors ${\bm r}_i^{(0)}$, i.e.
 ${\bm r}_i={\bm r}_i^{(0)}+\Delta{\bm r}_i$, where the displacement
vectors are normally distributed,
$\Delta{\bm r}_i \sim N_2(0,\sigma^2_{\Delta})$,  with zero expectation value
and variance, $\sigma^2_{\Delta}$.
This normal distribution of $\Delta{\bm r}_i$ mimics the anisotropy in the
shape  of real root systems, i.e. the fact that  the centres of mass of roots are displaced
with respect to  the seed (node) positions.
The toy models then differ
amongst each other by the form of the function used for
 $\rho_i({\bm r})$
 and therefore by the value of overlaps between two discs.

\begin{figure}[htp]
\includegraphics[angle=0,scale=0.5]{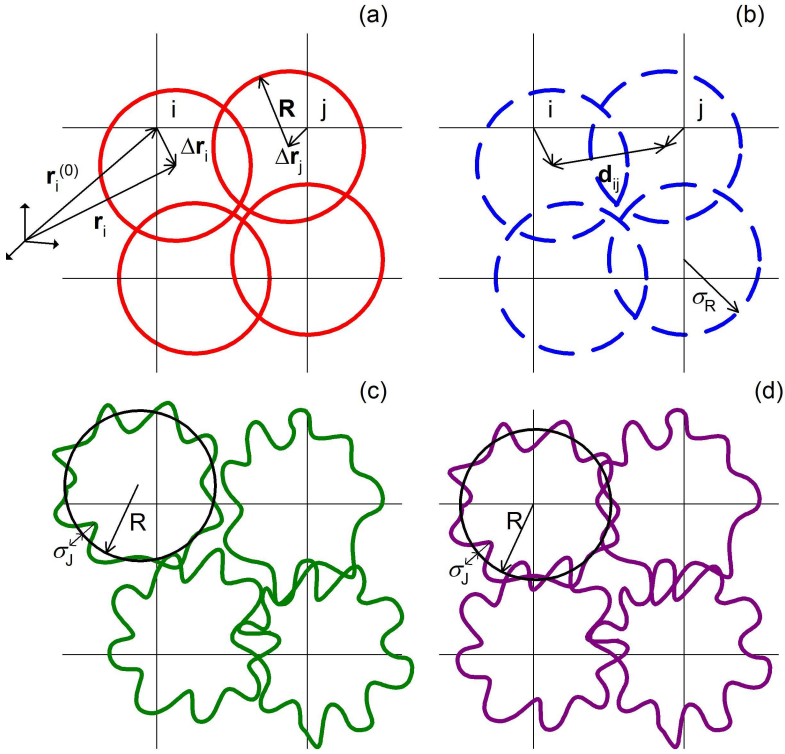}
\caption{\label{fig:discs}
Disc-based toy models: (a) displaced solid discs; (b) displaced
dispersed discs; (c) displaced discs with irregular edges (d)
non-displaced discs with irregular edges
}
\end{figure}

In the first model (see Fig.~\ref{fig:discs}(a)), all the discs
have the same radius, $R$, and  constant
number density, $\rho_i({\bm r})= \rho_0$ if
$|{\bm r} -{\bm r}_i|\le R $, and zero density
otherwise.
The overlap  between two discs $i$ and $j$
with centres separated by distance $d_{ij}=|{\bm r}_j-{\bm r}_i|$ reads
\begin{equation}
J^{(1)}_{ij} =
V_{\text{pixel}}\int \text{d}{\bm r} \rho_i({\bm r}) \rho_j({\bm r})
=
2\rho_0^2 R^2 \left[
\cos^{-1}(d_{ij}/2R) - (d_{ij}/2R) \sqrt{1-(d_{ij}/2R)^2}
\right]~,
\label{eq:J_1}
\end{equation}
if $d_{ij}/2R \le 1$ and zero otherwise (with the pixel area
$V_{\text{pixel}}=1$).
This is one of the simplest models, which is equivalent to
a representation of the real root systems by identical uniform vertical cylinders
projected onto  the horizontal plane.

%
%
%
%

In the second model (see Fig.~\ref{fig:discs}(b)), the density is
normally distributed around  ${\bm  r}_i$ with variance $\sigma^2_R$, i.e.
\[
\rho_i({\bm r})=
\tilde{\rho}(2\pi\sigma_R^2)^{-1}
\exp{\left(-|{\bm r} -{\bm r}_i|^2/2\sigma_R^2
\right)}~,
\]
where $\tilde{\rho}$ is a dimensionless normalization constant,
and the overlap between two dispersed discs represented by
Gaussian functions is
\begin{equation}
J^{(2)}_{ij} = \frac{{\tilde{\rho}}^2}{4\pi\sigma_R^2}
\exp{ \left(-d_{ij}^2/
    4\sigma_R^2\right)}
~.
\label{eq:J_2}
\end{equation}
The second model accounts for the property that the
number density in real roots is  a
decaying function with distance from the main tap root.

The third model (see Fig.~\ref{fig:discs}(c))  mimics the angular
irregularity  in the shape of real root systems,
i.e. the fact that overlap significantly depends on the
 angle $\varphi$ at which the roots  are placed at the lattice nodes
(see Fig.~\ref{fig:rootProj}).
In this model,
we assume that the overlap  has two
contributions,
\begin{equation}
J^{(3)}_{ij} = J^{(1)}_{ij} + \delta J_{ij}
~,
\label{eq:J_3}
\end{equation}
where $J^{(1)}_{ij}$ is given by Eq.~(\ref{eq:J_1}) as in the first
model and a random normally distributed
 constituent, $\delta J_{ij} \sim N(0,\sigma_J^2)$,
 with the variance, $\sigma_J^2$, which can depend on $J^{(1)}_{ij}$.

\begin{figure}[htp]
\includegraphics[angle=0,scale=0.6,clip=true]{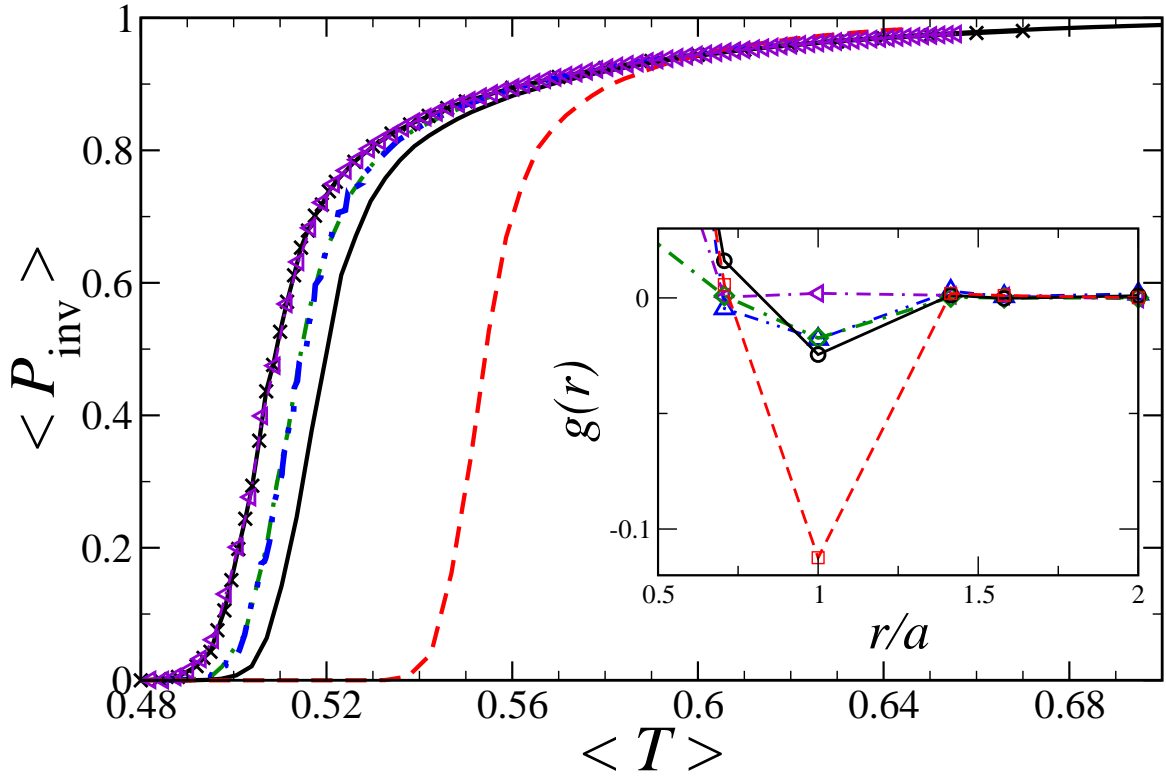}
\caption{\label{fig:modelTvsP}
The invasion curves for mean-field system (solid curve marked by the
crosses), which coincides exactly with that for the non-displaced irregular
discs model (dot-double dashed with triangles), real-root system
(solid), solid-disc model (dashed),
dispersed disc model (dot-dashed) and discs with irregular edges
(double dot-dashed).
The inset shows the correlation functions in the region around the
negative dip
(the same curve styles are used).
}
\end{figure}

In order to illustrate that irregularity in shape alone is
  insufficient to cause the observed effect, we concider a fourth
  model, which is a special case of the third model, where there is no
  displacment of the discs (i.e. we set $\sigma^2_{\Delta}=0$). Here
  we see no correlations (see the dot-double dashed line  with left
  triangles in the
  inset in Fig.~\ref{fig:modelTvsP}), and thus the system can be approximated by
  an equivilent homogeneous system (the dot-double dashed line  with left
  triangles coincides with the solid line marked by crosses
  in  in Fig.~\ref{fig:modelTvsP}).

The disc-based models depend on several parameters, such as
$R$,  $\sigma^2_\Delta$, $\sigma^2_R$ and $\sigma^2_J$.
The two normalization constants $\rho_0$ and  $\tilde{\rho}$ can be
incorporated into the value of the  infection efficiency $k$
and are set, for convenience, to $\rho_0=\tilde{\rho}=1$.
The values of the disc radius, $R$, in the first model and the width of the
Gaussians in the second model, $\sigma_R$, can be estimated from  the
mean value (amongst all $N_r$ roots) of the variance of the real root density
projected onto the horizontal plane, i.e.
\begin{equation}
R^2 = \sigma^2_R = N_r^{-1}\sum_i^{N_r} \delta {\bm r}_{\perp i}^2
~,
\label{eq:Radius}
\end{equation}
where
\begin{equation}
\delta {\bm r}_{\perp i}^2 = \int \text{d}{\bm r}
({\bm r}- {\bm r}_i)_\perp^2 \rho_i({\bm r})~, ~~~\text{and}~~~
{\bm r}_i=\int \text{d}{\bm r} ({\bm r}- {\bm r}_i^{(0)}) \rho_i({\bm r})
~,
\label{eq:Radius1}
\end{equation}
with $\rho_i({\bm r})$ given by Eq.~(\ref{eq:broaden}) and suffix
$\perp$ referring to the vector component perpendicular to the vertical
axis.
This value is found to be $R=\sigma_R \simeq 73$~px$^{1/2}$.
Similarly, the value of $\sigma^2_\Delta$ was estimated as
\begin{equation}
\sigma^2_\Delta =  N_r^{-1}\sum_i^{N_r} ({\bm r}_i- {\bm
  r}_i^{(0)})^2\simeq 1444 \text{px}
~,
\label{eq:sigma2}
\end{equation}
The value of $\sigma^2_J$ can be estimated from the variance of the
overlap  $J_{ij}(\varphi_i,\varphi_j)$
between two real roots which stochastically depends on their orientations
characterized by $\varphi_i$ and $\varphi_j$ (see Fig.~\ref{fig:rootProj}), and it
is found to be $\sigma_J/ \langle J^{(1)}_{ij}\rangle  \simeq
1.35$.

%
%
\vskip40pt
%
%
Having the estimates of all the parameters for the toy models it is
straightforward to evaluate numerically the dependence of the invasion
probability on mean transmissibility.
The results presented in Fig.~\ref{fig:modelTvsP} clearly demonstrate
that, for all the models exhibiting anisotropy, the critical point
$T_c$ is shifted to the right
of that for the mean field system
  and the invasion probability is also reduced in a rather wide
   range, $0\le P_{\text{inv}}\alt 0.95$, as compared with the homogeneous system
(note, that models 2 and 3  reproduce the real
root system  reasonably well).
This is a consequence of the short-range correlations in
transmissibilities present in the disc-based models.
The bigger shift for the first model as compared with less significant
shifts for models 2 and 3 reflects the respective decrease in the
scale of correlations for the latter models (cf. dashed, dot-dashed and
double dot-dashed
curves in Fig.~\ref{fig:modelTvsP} and corresponding curves
in the inset).
The relatively large correlations in the first model are due to sharp disc
boundaries so that the overlaps are significantly larger
(smaller)
for spatially close (distant) discs as compared with the mean value.
This difference is diminished by the Gaussian broadening of the disc
density so that the correlations for the second model are less
pronounced.
The angular fluctuations of density (model 3) result in a similar
reduction of correlations found for the solid disc model.

We conclude from the analysis of the toy models that anisotropy
in the  root shape
accounts for most of the short-range correlations in transmissibilities between
the root systems. Anisotropy arises mainly
 from  the horizontal shift of the centre
of mass with respect to the lattice node (seed) location. To the
greatest extent, this effect is  picked up by the first  model
 which displays the most significant effect of correlations.
Dispersion and irregularity in the root shapes mimicked by models 2 and 3,
respectively, reduce the level of correlations.
 Finally, if irregular discs are not displaced from the lattice nodes
(model 4)
the correlations completely disappear and the model is equivalent to
the mean-field one. 

\section{Discussion  \label{sec:discussion}}

The occurrence of well-defined phase transitions in determining
whether or not a pathogen invades through a lattice of homogeneous
hosts is well established \citep{Grassberger_83}.
 Following initial work by \cite{Bailey_2000}
 on the spread of the soil-borne fungus, \emph{R. solani}, critical
percolation phenomena and phase transitions in lattice-based systems
have been supported by experimental analysis. More recently, the concept
has been extended to the transmission of soil-borne infections through
lattices with missing sites \citep{Otten_2004}
and through animal populations, exemplified by the spread of plague
through gerbil populations in lattices of interconnecting burrows
\citep{Davis_2008}.
Here we have extended the theoretical analysis of
critical percolation phenomena to take account of the spatial correlations
that occur in the set of morphologically complex 3d hosts
  represented by plant root systems.

We have used digitized images of the root systems of bean seedlings
to provide realistic examples of a complex morphological branching
structure to test the methodology.
However, our results and methodological approaches, in particular the new disc-based models,
provide a basis for further analysis of the transmission of infection and disease
through a broad range of morphological structures.

We have shown that the invasion probability $(P_{inv})$ of a
  root-infecting pathogen spreading through a population of plants
  with morphologically-complex, overlapping 3d root systems,
  depends upon
  the transmissibility of infection between neighbouring plants
  (Eq.~(\ref{eq:transmissibility})).  We have resolved the
  transmission into two components, the transmission efficiency $(k)$
  between nearest neighbours, and the lattice spacing between plants
  $(a)$, which control the degree of overlap.  The transmission
  efficiency is a measure of the probability of infection  arising
  from overlap between an infected and a susceptible root system. Here
  we assume a constant transmission rate per overlapping voxel for
  overlapping roots but this assumption could be relaxed.  The extent
  of the overlap is broadened to allow for the ability of the pathogen
  to explore soil immediately surrounding a root, known as the
  pathozone \citep{Gilligan_1983}. The spacing between plants controls the
  degree of overlap. We have used the mapping to bond percolation in
  order to calculate the invasion probability for a set of
  transmissibilities, showing evidence for a marked phase transition
  that depends upon $k$ and $a$ (Fig.~\ref{fig:3d_phase_diagram}). We
  conclude that the heterogeneities in transmissibilities
  arising from correlations in degree of overlap between neighbouring
  plants, renders a population of root systems more
  less vulnerable 
 to
  epidemic invasion than a corresponding uncorrelated homogeneous system.

\begin{figure}[htp]
\includegraphics[angle=0,scale=0.6,clip=true]{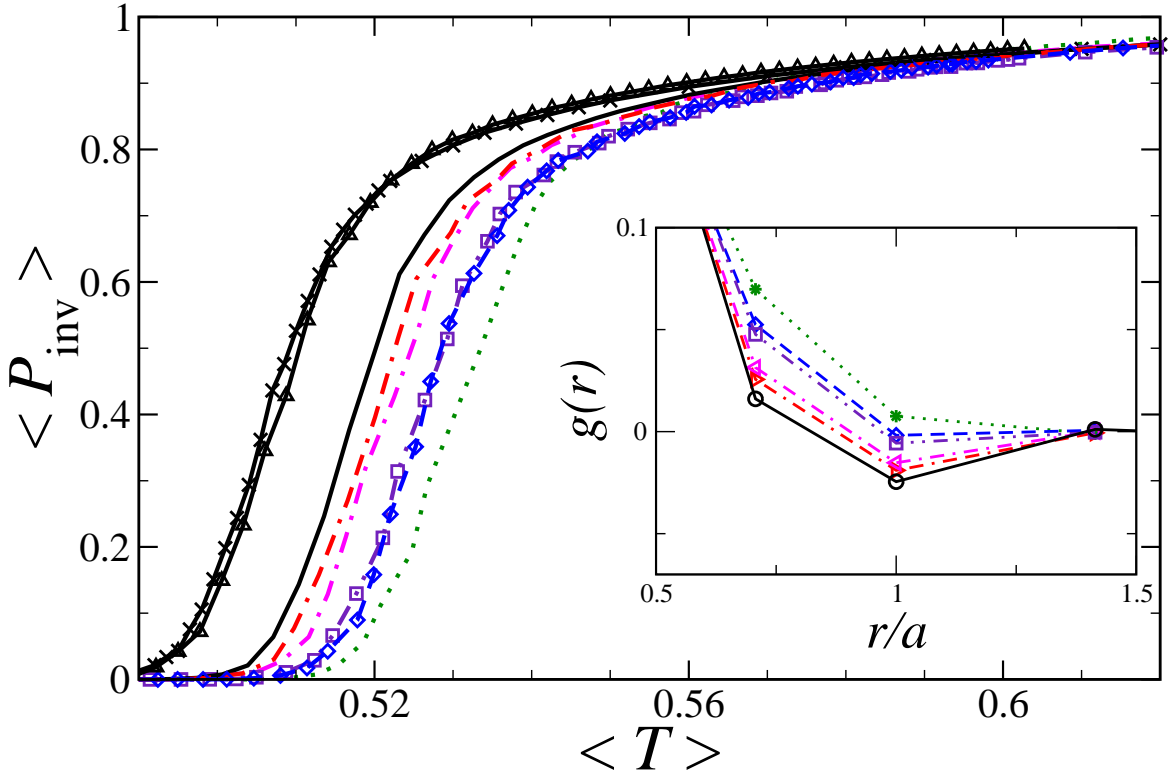}
\caption{\label{fig:disorder_in_tau}The invasion curves for
    the mean-field system (solid curve marked by the
crosses), the real-root system with homogeneous recovery times (solid),
and distributed with $\text{log}(\tau)\sim N(0,0.5)$ (dot
double-dashed),
$\text{log}(\tau)\sim N(0,1)$ (dashed with diamonds),
$\tau\sim N(1,0.5)$ (dot-dashed),
$\tau\sim N(1,1)$ (double dot-dashed with squares),
$\tau\sim \text{Exponential}(1)$ (dotted). 
 The solid line marked by triangles represents an 
 anisotropic system of
identical roots in identical orientation.   
The inset shows the correlation functions (the same curve styles are
used).
}
\end{figure}

For concreteness, we have analyzed the square-lattice topology
only. Any other regular lattice topologies can be treated in a similar
way.
The invasion curve, $P_{\text{inv}}(\langle T \rangle)$, is specific for
the lattice topology but the effect of correlations in
transmissibilities between root systems will be qualitatively the same as
that found for the square lattice.
In modelling the spread of
  an SIR epidemic through heterogeneous, overlapping root systems, we
have used several simplifications.
One of these is related to the
assumption of identical recovery times for all the hosts. If this
assumption is relaxed and the recovery times are distributed according
to a certain probability density function then
 the invasion probability in
such a system is lower than that in the
system with homogeneous recovery times
 \citep{Kuulasmaa_82,Cox_88}.
 We have demonstrated this using several realistic distributions
   for recovery times in root systems.
The results presented in Fig.~\ref{fig:disorder_in_tau} clearly show a
shift of invasion curves to the right, thus indicating that
systems with heterogeneity in recovery times are indeed more
resistant 
 to SIR
epidemics.
The main reason for such increased resistance is that
the  heterogeneity in recovery times brings positive
correlations to the system (see inset in
Fig.~\ref{fig:disorder_in_tau} and
~\cite{Kuulasmaa_82,Cox_88,Miller2007,Kenah_07}).

%
%
\vskip40pt
%
%

Another assumption in our approach is that the root
anisotropy is in a random direction for each root.
However, in a real planting, large scale resource gradients can lead
to a predominant orientation of all root systems.
This results in anisotropy of the overlaps $J_{ij}$  meaning that the
root overlaps along a certain direction are greater than those in other
directions. 
In fact, the system becomes more ordered typically exhibiting larger transmissibilities in
the direction of resource gradients. 
For illustration, we considered an extreme case for which an epidemic
spreads 
through a system of anisotropically identical roots placed in
the same orientation (without rotation) on all the nodes of a square
lattice. 
The results are presented by the solid curve marked by triangles 
 in Fig.~\ref{fig:disorder_in_tau} from
which it follows that the system of identical anisotropic roots is
more vulnerable than the system of randomly rotated roots (cf. the
dashed curve with the solid one) meaning that the external gradients
may reduce the resilience of the system. 
This is an expected effect because the SIR epidemic in the system of
identical anisotropic roots can be mapped onto anisotropic percolation
characterized by two transmissibilities $T_x \ne T_y$ along different
lattice directions. 
It is known \citep{Sykes_Essam1964}, that the invasion threshold in
such a system coincides with that in homogeneous one, i.e. given by
the following equation, 
$T_c=\langle T \rangle =(T_x+T_y)/2 =1/2$ (for square lattice), which
can be seen from Fig.~\ref{fig:disorder_in_tau} (cf. solid lines marked
by triangles and crosses). 

In the analysis above we have used the toy models in order to
  understand the spread of epidemics in systems of morphologically
  realistic roots.
The models were designed in such a way as to capture the main
features of root systems that influence the spread of epidemics.
Our analysis of the toy models reveals that anisotropy,
  rather than disorder  in root shape, 
has the major effect on epidemic invasion.
The anisotropy in the real root systems is 
due to the fact that the 3d
  root density distribution is not centrally located beneath the seed.
By mimicking a similar effect within toy models, we find that
the greater the degree of root anisotropy, the
  more resistant 
 the system is to epidemic invasion.
The
effect is due to anisotropy-induced, short-range correlations in
disease transmissibilities between different root systems which result in
a reduced probability of invasion.
These correlations are such that a root system that is well connected with one
of its neighbours is likely to be poorly connected with its neighbour
on the opposite side.

The other root characteristic, morphological complexity, has an
opposite effect on the probability of epidemic invasion.
By morphological complexity, we mean that the roots branch producing
secondary roots so that the shape of the root system becomes irregular.
The increase in degree of branching makes the system more isotropic
and diminishes the correlation effects thus enhancing the vulnerability
of the system to epidemics.
This has been demonstrated with toy models by introducing
  disorder, characterised
  by the parameter $\sigma_J$ mimicking the degree of branching.
 Overall we can conclude that highly
anisotropic and poorly branching root systems
render the population less prevalent to invasion of soil-borne pathogens.

In our analysis, we have used the roots of young bean seedlings, grown
under hydroponic conditions.
These serve to illustrate the  methodology.
The hydroponically grown bean roots exhibit an inherent anisotropy in
their shapes which is the main morphological factor in reducing the
invasion probability. Further work will consider the root morphology
of monocotyledenous as well as dicotyledenous root systems grown and
imaged in soil.
We also assume quenched disorder in the systems
considered here. This means that all the properties of the hosts,
including their morphology, do not depend on time. Such an assumption
is true only in the case of relatively fast epidemics and/or slowly
growing root systems. Of course, in real situations, the root
morphology can significantly change in the course of an epidemic
requiring time-dependent transmissibilities, that are beyond the
  scope of this paper but comprise an important topic for further
  study.

\section{Acknowledgments} TPH would like to thank the UK EPSRC
  for
financial support. FJPR, SNT, FMN and CAG thank BBSRC for funding (Grant
  No. BB/E017312/1). Luciano da F. Costa thanks CNPq
(301303/2006-1 and 573583/2008-0) and FAPESP (05/00587-5) for
sponsorship. Mauro Miazaki thanks FAPESP (07/50988-1) for his grant
and Christopher Gilligan gratefully acknowledges the support of a
BBSRC Professorial Fellowship.
The authors thank Alexandre Cristino for suggestions regarding the
growth of the seeds.

\bibliographystyle{JRSstyle}

\end{document}